\documentclass[a4paper,11pt]{article}
\pdfoutput=1 

\usepackage{jinstpub} 

\usepackage{gensymb}

\title{\boldmath A New Beam Polarimeter at COSY to Search for Electric Dipole Moments of Charged Particles}

\author[1,2]{F.~M\"uller,}
\author[1,3]{O.~Javakhishvili,}
\author[4]{D.~Shergelashvili,}
\author[1,*]{I.~Keshelashvili,}
\author[4]{D.~Mchedlishvili,}
\author[1,2]{F.~Abusaif,}
\author[5]{A.~Aggarwal,}
\author[6]{L.~Barion,}
\author[6]{S.~Basile,}
\author[1]{J.~B\"oker,}
\author[6]{N.~Canale,}
\author[6]{G.~Ciullo,}
\author[6,7]{S.~Dymov,}
\author[1]{O.~Felden,}
\author[3]{M.~Gagoshidze,}
\author[1]{R.~Gebel,}
\author[1]{N.~Demary,}
\author[1]{K.~Grigoryev,}
\author[1]{D.~Grzonka,}
\author[1]{T.~Hahnraths,}
\author[1]{V.~Hejny,}
\author[1]{A.~Kacharava,}
\author[1]{V.~Kamerdzhiev,}
\author[5]{S.~Karanth,}
\author[7]{A.~Kulikov,}
\author[1,2,8]{A.~Lehrach,}
\author[6]{P.~Lenisa,}
\author[4]{N.~Lomidze,}
\author[9]{B.~Lorentz,}
\author[4]{G.~Macharashvili,}
\author[5]{A.~Magiera,}
\author[3]{Z.~Metreveli,}
\author[1]{A.~Nass,}
\author[10,11]{N.N.~Nikolaev,}
\author[4]{M.~Nioradze,}
\author[1,6]{A.~Pesce,}
\author[1]{V.~Poncza,}
\author[1]{D.~Prasuhn,}
\author[1,2,8]{J.~Pretz,}
\author[1]{F.~Rathmann,}
\author[1]{A.~Saleev,}
\author[1]{T.~Sefzick,}
\author[11,12]{Yu.~Senichev,}
\author[1,7]{V.~Shmakova,}
\author[2]{J.~Slim,}
\author[13]{H.~Soltner,}
\author[14]{E.~Stephenson,}
\author[1,8]{H.~Str\"oher,}
\author[4]{M.~Tabidze,}
\author[15]{G.~Tagliente,}
\author[7,16,17]{Yu.~Uzikov,}
\author[1]{Yu.~Valdau,}
\author[1,2]{T.~Wagner,}
\author[5]{A.~Wro\'{n}ska,} 
\author[13]{P.~W\"ustner,}
\author[1]{and M.~\.{Z}urek}
\affiliation[1]{Institut f\"ur Kernphysik, Forschungszentrum J\"ulich, 52425 J\"ulich, Germany}
\affiliation[2]{III. Physikalisches Institut B, RWTH Aachen University, 52056 Aachen, Germany}
\affiliation[3]{Dept.of Electrical and Computer Engineering, Agricultural University of Georgia, 0159 Tbilisi, Georgia}
\affiliation[4]{High Energy Physics Institute, Tbilisi State University, 0186 Tbilisi, Georgia}
\affiliation[5]{Marian Smoluchowski Institute of Physics, Jagiellonian University, 30348 Cracow, Poland}
\affiliation[6]{University of Ferrara and INFN, 44122 Ferrara, Italy}
\affiliation[7]{Laboratory of Nuclear Problems, Joint Institute for Nuclear Research, 141980 Dubna, Russia}
\affiliation[8]{JARA--FAME (Forces and Matter Experiments), Forschungszentrum J\"ulich and RWTH Aachen University, Germany}
\affiliation[9]{GSI Helmholtzzentrum f\"ur Schwerionenforschung, GmbH, 64291 Darmstadt, Germany}
\affiliation[10]{L.D. Landau Institute for Theoretical Physics, 142432 Chernogolovka, Russia}
\affiliation[11]{Moscow Institute for Physics and Technology, 141700 Dolgoprudny, Russia}
\affiliation[12]{Institute for Nuclear Research of the Russian Academy of Science, 117312 Moscow, Russia}
\affiliation[13]{Zentralinstitut f\"ur Engineering, Elektronik und Analytik, Forschungszentrum J\"ulich, 52425 J\"ulich, Germany}
\affiliation[14]{Indiana University Center for Spacetime Symmetries, Bloomington,  Indiana 47405, USA}
\affiliation[15]{Istituto Nazionale di Fisica Nucleare sez. Bari, 70126 Bari, Italy}
\affiliation[16]{Dubna State University, 141980 Dubna, Russia}
\affiliation[17]{Department of Physics, M.V. Lomonosov Moscow State University,119991 Moscow, Russia}

\affiliation[*]{Corresponding author.}

\emailAdd{i.keshelashvili@fz-juelich.de}

\abstract{
A calorimetric polarimeter based on inorganic LYSO scintillators is described. 
It has been designed for use in a storage ring to search for electric dipole moments (EDM) of charged particles such as the proton and deuteron. 
Its development and first use was on the Cooler Synchrotron (COSY) at the
Forschungszentrum J\"ulich with 0.97~GeV/c polarized deuterons, a particle and energy suitable for an EDM search. 
The search requires a polarimeter with high efficiency, large analyzing power, and stable operating characteristics. 
With typical beam momenta of about 1~GeV/c, the scattering of protons or deuterons from a carbon target into forward angles becomes a nearly optimal choice of an analyzing reaction. 
The polarimeter described here consists of 52 LYSO detector modules, arranged in 4 symmetric blocks (up, down, left, right) for energy determination behind plastic scintillators for particle identification via energy loss.
The commissioning results of the current setup demonstrate that the polarimeter is ready to be employed in a first direct search for an EDM on the deuteron, which is planned at COSY in the next two years.
}


\collaboration{\includegraphics[height=17mm]{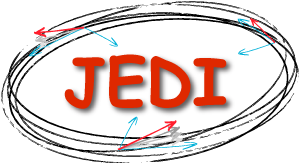}\\[6pt]
   JEDI collaboration}

\begin{document}

\maketitle

\flushbottom


%
\section{Introduction}
\label{Intro}

This paper describes a spin polarimeter for intermediate energy proton and deuteron beams in a storage ring that is designed specifically to search for an electric dipole moment (EDM) on the beam particles. See \cite{Abusaif} for a project description and \cite{Farley,Yannis} for general descriptions of the technique. This work follows similar suggestions for using electric fields to create a measurable torque on the EDM of stored particles \cite{Nelson,Kriplovich}. In this paper, the effect of the torque is observed by measuring the change over time in the asymmetry of beam particles scattered through the spin-orbit interaction with a spin-0 carbon target.

It has been proposed that a storage ring might be used to extend the sensitivity of searches for an EDM by exploiting the realm of charged particle 
beams
for which good statistical data are obtainable from highly efficient polarimeters
\cite{Farley,Yannis}. 
The design of the storage ring must 
ensure 
the correct balance between electric and magnetic bending in the arcs so that in the ring plane (in-plane) the beam polarization direction and the particle velocity rotate at the same rate, a property 
called “frozen spin.”
For the proton, the choice of momentum $p = 700.7$~MeV/c 
permits the use of only 
electric fields for bending the particle beam 
and the chance to cancel systematic errors by using simultaneously counter-rotating ($CW$ and $CCW$) beams. 
With the polarization direction 
maintained 
by feedback to be always parallel to the 
velocity vector,
 the particle frame radial electric field exerts a torque 
via
the EDM that slowly rotates the polarization out of the ring plane and 
towards
the vertical direction, thus creating the EDM signal 
via
a changing left-right scattering asymmetry from a target that is a part of an in-beam polarimeter.
This paper reports on 
a prototype 
of 
such a polarimeter that has been installed and tested on the Cooler Synchrotron (COSY) at 
the 
Forschungszentrum J\"ulich \cite{Maier} within the framework of the JEDI (J\"ulich Electric Dipole moment Investigations) Collaboration \cite{JEDI}. 
This experiment is one of a series whose purpose is to demonstrate the feasibility of high-sensitivity polarization measurements \cite{Brantjes}, long-lifetime in-plane polarization \cite{Guidoboni}, and real-time feedback control of the polarization direction \cite{Hempelmann}.
This is a subset of the
requirements 
to show
the feasibility of such a storage ring project. 

With the addition of an RF Wien filter 
\cite{SlimNIM, SlimJINST}
to the COSY ring to break the cancellation of EDM effects in the absence of frozen spin, an experimental program 
has been started
to place an upper limit on the EDM of the deuteron \cite{Rathmann}. 
Recently it has also been suggested that the COSY ring with a stored, horizontally-polarized beam (not frozen spin) 
could,
by ramping the energy of the machine, be used to search for axion-like particles by crossing a resonance at the frequency associated with the axion mass \cite{Chang, Pretz}. 
In 
this
case 
a jump in the vertical component of the polarization as the machine crosses the 
resonance should be observed. 
All of these experiments involve the combination of a storage ring with a horizontally polarized beam and a 
high-efficiency 
polarimeter to track that polarization throughout a beam store where signals of new physics are expected to appear.

The polarimeter features a segmented calorimeter consisting of LYSO crystals read out using 
silicon photomultipliers (SiPMs)
\cite{Irakli-1, FabianPhD}
immediately behind a thin plastic scintillator (also with SiPM readout) used for particle identification. 
These detectors observe either proton or deuteron scattering from a thick carbon target at intermediate energies. 
The entire assembly fits into a 1.3~m space in a COSY ring straight section.

The next section presents in more detail the requirements for a polarimeter that would be suitable for an EDM search using either proton or deuteron beams in a storage ring. 
This is followed by sections that provide more information concerning the setup, detector construction and data acquisition, polarimeter signal characteristics, COSY installation, and initial commissioning results. 
The last section discusses the 
current status of the project
and prospects for continued development.

%
\subsection{General Considerations}
\label{GenCons}

A 
high-sensitivity 
polarimeter, driven by the desire for the best possible search, requires that small statistical errors from 
very 
high event totals 
can be obtained within a reasonable running time. 
In addition, the analyzing power, which sets the scale of the sensitivity of the scattering to beam polarization, should be as large as possible. 
This leads to the choice of intermediate beam energies (momenta about 1~GeV/c), robust targets in the neighborhood of carbon or somewhat higher mass, and 
forward-angle 
elastic scattering as the ideal polarimeter concept. 
Useable scattering angles may range from the outer edge of the Coulomb-nuclear interference at a few degrees to roughly $\theta_{lab} = 15^\circ$. 
These features are almost independent of nuclear structure, arising mainly from the central and spin-orbit interactions between the projectile and the nucleus that are well described by the optical model \cite{Bechetti,Daehnick,Meyer-200,Meyer-250,Kawabata,Satou}. 
This makes the isolation of elastic scattering from all other reactions less critical in the design,
since inelastic scattering and even particle transfer reactions with low reaction Q-values show similar analyzing powers. 
The major exception applies to reactions producing protons from deuteron breakup,
which are almost spin independent. 
If possible, these protons should be removed from the flux going toward the polarimeter before they reach any active detector elements. 
In a remarkable coincidence, the best place to operate the proton EDM search is at $p = 700.7$~MeV/c ($T_p = 232.8$~MeV) where the analyzing power for a $p + C$ polarimeter is near its maximum ($T_p = 210$~MeV \cite{McNaughton}).

\begin{figure}[hbtp] \centering
    \includegraphics[width=4cm]{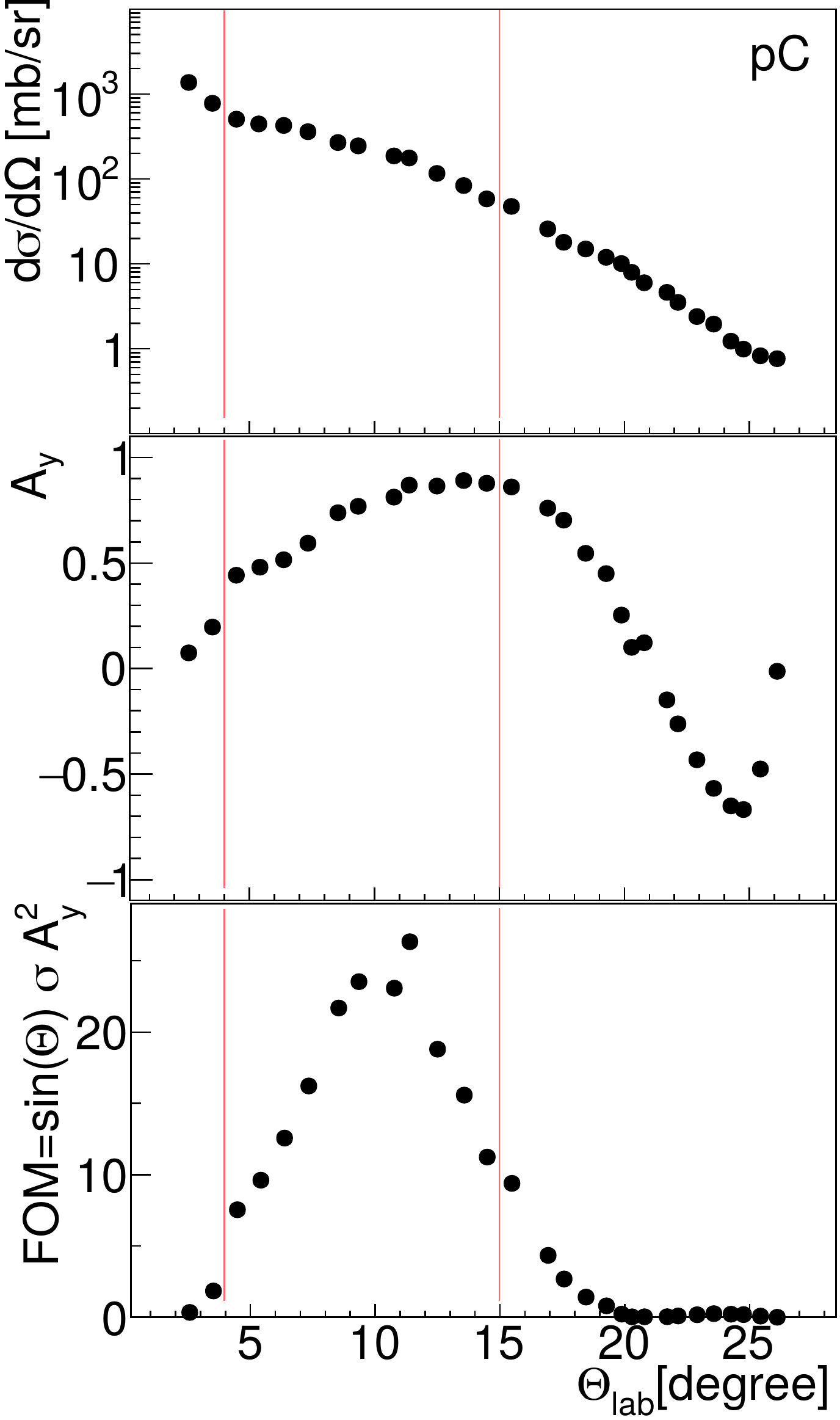}
    \hspace{2cm}
    \includegraphics[width=4cm]{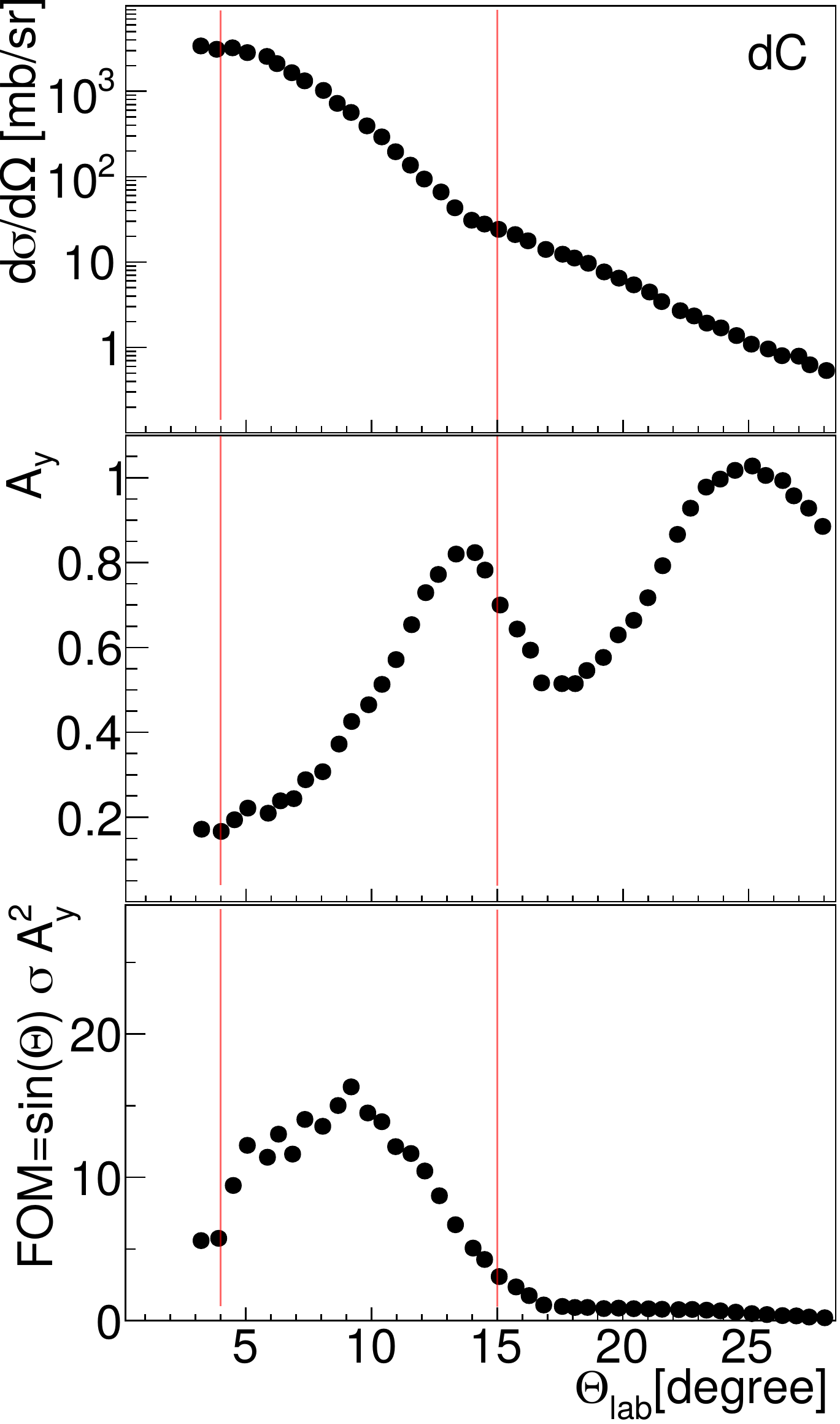}
    \caption{Measurements of the laboratory frame differential cross section, analyzing power, and modified figure of merit [$\sigma (\theta ) A_{y}^{2} (\theta ) sin\theta$] for proton scattering from carbon at $T_p = 250$~MeV \cite{Meyer-250} (left) and deuteron scattering from carbon at $T_d = 270$~MeV \cite{Satou} (right).
}
    \label{figFOM}
\end{figure}

To illustrate the selection of appropriate operating angles for the polarimeter detectors, consider the laboratory frame cross sections and analyzing powers 
for the scattering reactions
$p + C$ and $d + C$ in Fig. \ref{figFOM}.
The figure also shows a modified figure of merit. 
The usual definition, $FOM=\sigma (\theta ) A_{y}^{2}(\theta )$ where $\sigma (\theta )$ is the differential cross section and $A_y (\theta)$ is the vector analyzing power, has been augmented with $sin(\theta_{\rm lab})$ to recognize the increase in the available solid angle for all azimuthal angles as the polar scattering angle increases.
This figure of merit varies as the inverse square of the statistical error in a polarization measurement, hence it is important to maximize the integral of this quantity in the choice of polarimeter angle coverage. 
At the same time, areas of low analyzing power should be avoided,
since a large analyzing power provides more leverage against systematic errors. 
Red lines ($4^\circ$ to $15^\circ$ in Fig. \ref{figFOM}) mark a range that would be desirable to exploit in a polarimeter detector scheme.

\begin{figure}[hbtp] \centering
    \includegraphics[height=8cm]{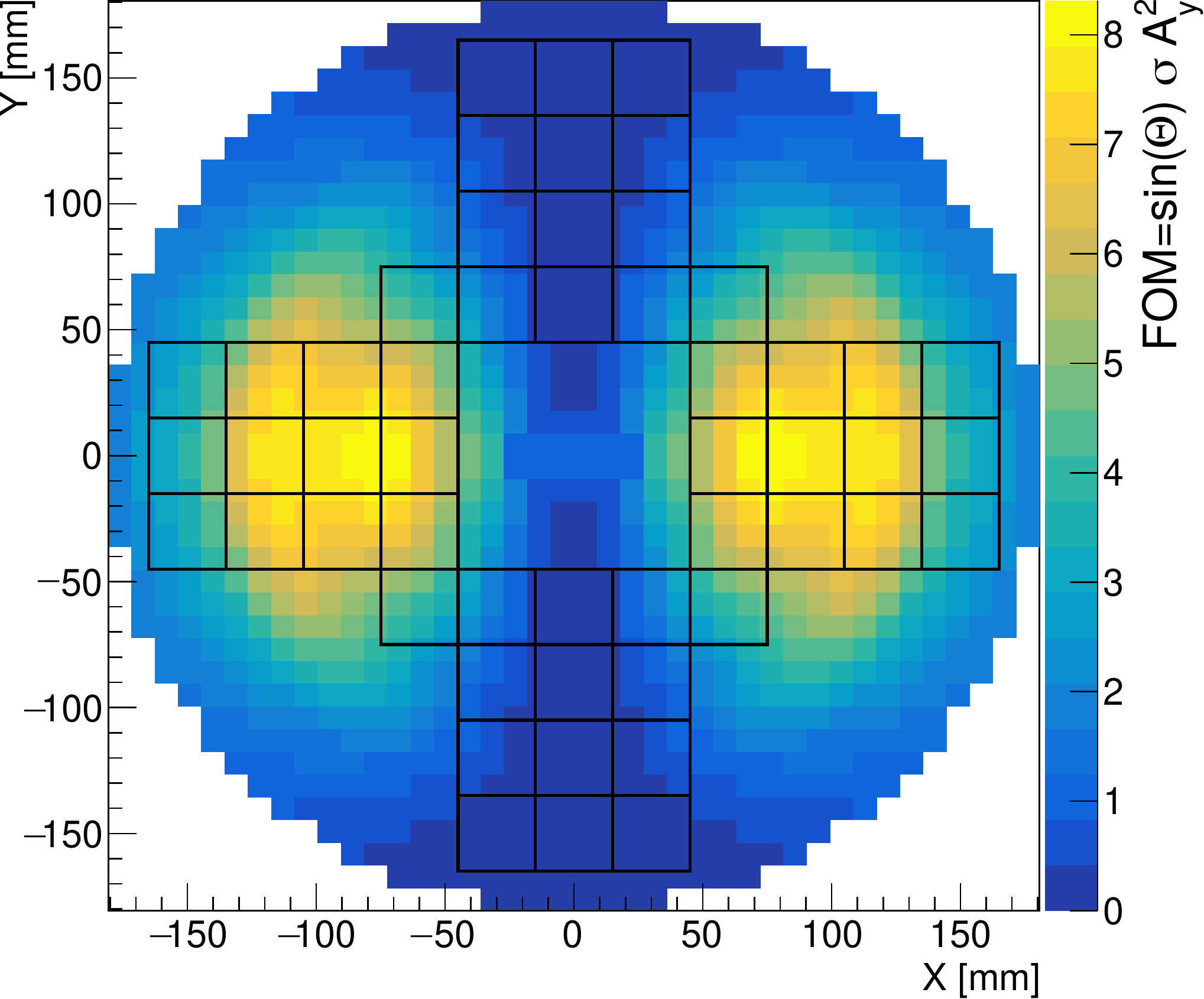}
    \caption{Variation by color of the figure of merit across the polarimeter based on $d + C$ elastic scattering \cite{Satou}. The black lines show the locations of the 52 LYSO crystals used in the current version of the EDM polarimeter.}
    \label{figSimFOM}
\end{figure}

Figure \ref{figSimFOM} shows a 2-D map of the figure of merit across the polarimeter front face for a vertically polarized beam. 
The best areas (yellow) are to the left and right of the beam center. 
The figure also includes the dependence on the cosine of the azimuthal angle ($\phi$), 
which
appears in the formula for the differential cross section dependence on spin \cite{Tanifugi} for protons:
\begin{equation}
    \sigma_{\rm pol}(\theta,\phi)=\sigma_{\rm unpol}(\theta)[1+p_yA_y(\theta)cos\phi]
    \label{eq1}
\end{equation}
where
$p_y$ is the vertical component of the polarization.
For deuterons with spin-1, an additional factor of 1.5 appears ahead of the $pA$ term.
The rectangular grid superimposed on Fig. \ref{figSimFOM} illustrates the front-face locations of the presently installed (52) LYSO calorimeter modules. 
This grid overlaps strongly the 
most useful region in terms of
polarimeter statistics. 
This 2-D pattern applies to the asymmetry generated by Eq. \ref{eq1}, which, for example, produces a left-right 
($L/R$)
difference in the scattering rates for an upwardly polarized projectile. 
This pattern may be rotated by 90$^\circ$ to correspond to an up-down ($U/D$) asymmetry associated with the measurement of a sideways polarization. 
Both pieces of information are important. 
In the EDM search, the EDM signal appears as a slowly 
increasing
vertical component ($L/R$ asymmetry). 
The in-plane polarization direction will be controlled using real-time feedback to the storage ring RF cavity so that the polarization is always parallel to the particle velocity. 
This condition will be checked, and corrected, based on the deviations of the $U/D$ asymmetry from zero. 
The axion experiment \cite{Chang, Pretz}, which searches for an oscillating EDM, will look for a jump in the $L/R$ asymmetry as the storage ring revolution frequency is scanned. 
The $U/D$ signal will be monitored after unfolding the in-plane rotation \cite{Guidoboni, Bagdasarian} to verify that polarization is not lost due to decoherence of the particle spin tunes.

In order to reach the sensitivity goal of the EDM search of 10$^{-29}$~e$\cdot$cm on the proton, the polarimeter must be able to detect rotations into the vertical plane at the level of $\mu$rad using beam storage times of at least 1000~s \cite{Yannis} 
accumulated
during a total operating time of 
about one year. 
This requires more than 10$^{12}$ good polarimeter events. 
The beams will consist of 10$^{10}$ particles per fill, split between counter-rotating beams. 
This implies trigger rates before event selection in the range of 10$^6$~/s, which must be recorded with minimal rate-dependent distortions. 
This suggests that the polarimeter would work best in trigger mode, responding to information above some threshold rather than processing pulse height or timing information in order to make a more intricate cut. 
Stability will be important so as not to distort any changes in the vertical polarization between the beginning and the end of a beam store. 
Thus, any cuts that define events should be insensitive to changes in rates or the machine environment. Final analysis of such data will likely involve corrections for systematic rate and geometry errors \cite{Brantjes}.

In many polarization measurements, there are certain common strategies that are adopted to minimize systematic errors. 
Asymmetric count rates measured with two detectors ($L/R$ or $U/D$) eliminate problems associated with changes in the primary beam intensity. 
Two
opposite vector polarization states with detectors to the left and right, which together generate four counting rates, can be combined 
to define
a “cross-ratio” scheme that cancels many first-order errors \cite{Ohlsen}, using
\begin{equation}
    pA=\frac{r-1}{r+1}
    \hspace{2cm}
    r^2=\frac{L_+R_-}{L_-R_+}\,\,,
    \label{eqCR}
\end{equation}
where $L$ and $R$ represent event rates for two sides of the beam, and + and – indicate the polarization state of the beam. 
These errors often arise from a broken geometric symmetry caused by position or angle shifts in the beam or differences in opposite detector responses. 
Studies conducted earlier 
at
COSY \cite{Brantjes} show that it is possible to calibrate the changes in the asymmetry measurements when geometric or 
rate-related 
errors arise. 
The choice of detecting forward-going reaction products allows a single systematic driving term
\begin{equation}
    \phi=\frac{s-1}{s+1}
    \hspace{2cm}
    s^2=\frac{L_+L_-}{R_+R_-}
\end{equation}
to be used to estimate the corrections needed for both position and angle errors. 
Tests \cite{Brantjes} showed that the corrections cancel errors to below 10$^{-5}$ of the signal and show no
problems beyond that level. 
Such corrections are required for a successful EDM search.

The addition of unpolarized storage ring fills may provide additional leverage against systematic errors. 
If the $L$ and $R$ rates in Eqs.~1.2 and 1.3 for one polarization state are replaced with the $L_0$ and $R_0$ rates for an unpolarized measurement and the square of the second equation in each case is removed, then the formulas provide an alternative measurement of $pA$ for the remaining state by itself. These new equations are known as half cross ratios as shown by
\begin{equation}
    pA=\frac{r-1}{r+1}
    \hspace{2cm}
    r=\frac{L_+R_0}{L_0R_+}\,\,.
    \label{eqHCR}
\end{equation}

The carbon block target used 
during
the commissioning tests was 20~mm thick. 
The deuterons used for the commissioning run had $p = 970$~MeV/c ($T_d = 238$~MeV). 
Given the angle coverage shown in Fig. \ref{figSimFOM}, the efficiency of 
detector arrays should be about 1~\%. 
This efficiency is defined 
as
the ratio of the detected events used for a polarization measurement to the number of stored beam particles lost in the process. 
This is close to the limit for any polarimeter configuration \cite{Bonin-1, Bonin-2, Ladygin}.

Several design choices were made in the development of this polarimeter. 
Some examples from which to gain design ideas include devices intended to operate at the focal planes of magnetic spectrometers \cite{McNaughton}, where the composition of the particle flux is limited by the acceptance of the magnet system. 
In the cases 
of
protons, particle tagging without a calorimeter measurement was sufficient to select useful events. 
However, if a
polarimeter
operates for deuterons, then particle identification is critical. 

The carbon block has demonstrated that such a target may be operated near a stored beam and that 
results 
in 
a very efficient use of the beam particles in polarization measurements \cite{Brantjes}. 
However, it has the disadvantage that sampling of the beam is not uniform. 
Electron cooling, which is usually done for typical times of 75~s prior to any use of the beam, has the potential to homogenize the beam, thus reducing the chance of a significant polarization 
variation over the beam cross section.
However, earlier tests at higher intensities have shown that the beam exhibits features that are consistent with there being some segregation between outer halo parts, where the polarization lifetime is smaller and a core, where it is larger. 
At present, work has started on a pellet system that sends small targets (e.g. diamond pellets) through the beam while
tracking where they go \cite{JKM}. 
Then, a ray tracing detector system may be used, thus providing information to unravel the polarization from different parts of the beam profile. 
No results from such a system will be reported here.

\section{Detectors and Readout}
\label{secDetDAQ}

%
\subsection{General Plan}
\label{secGenPlan}

\begin{figure}[hbtp] \centering
    \includegraphics[height=6cm]{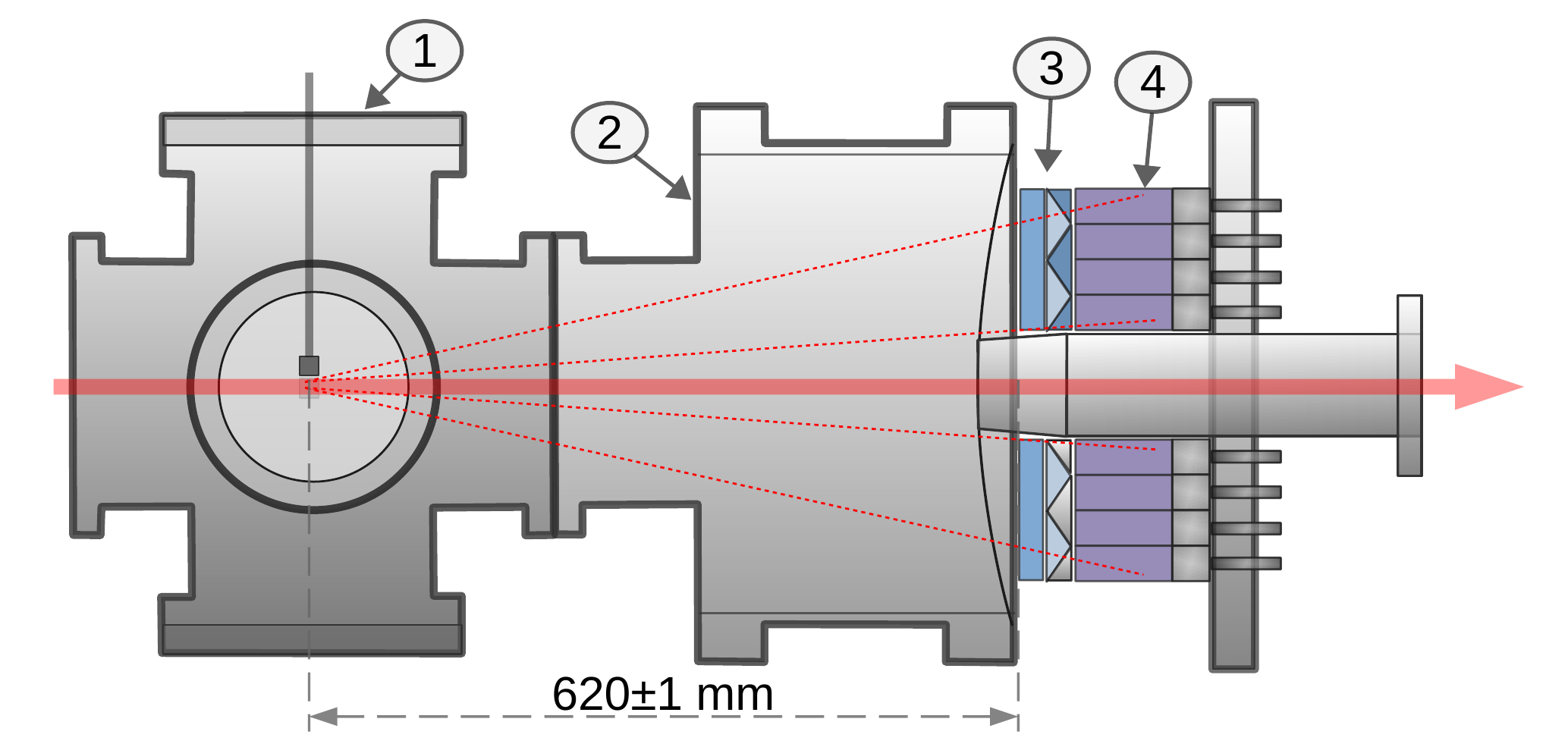}
    \caption{Polarimeter conceptual design (see text).
    (1) A target chamber equipped with movable vertical and horizontal 2~cm thick carbon blocks.
    (2) A vacuum flight chamber with 800~\micro m stainless steel exit window.
    (3) 2+2~cm thick plastic scintillator wall as tracking and $\Delta E$ counter.
    (4) The LYSO modules where the scattered projectiles are stopped.}
    \label{figJEPOSketch}
\end{figure}

A conceptual diagram of the main parts of the EDM polarimeter is shown in Fig. \ref{figJEPOSketch}.

The beam proceeds from left to right, passing by a carbon block target suspended from above by a movable support rod. 
The rod 
can 
be moved remotely and is used to remove the target during beam injection and acceleration. 
The beam-side target edge 
is
placed on the beam centerline to reduce systematic errors in the polarimeter asymmetry due to any displacement from the center. 
It is now current practice at COSY to add a beam bump to displace the beam at the target position by about 3~mm while running for data acquisition. 
Excitation of betatron oscillations using a white noise source connected to upstream vertical strip line plates brings particles from the beam to the front face of the carbon block \cite{Brantjes}.

The target chamber expands from the target location going downstream until it reaches the exit
window, which is made of 800~\micro m stainless steel. 
Once in air, scattered beam particle next encounter a plastic scintillator energy-loss detector.


Lastly, the particles enter heavy inorganic (LYSO, \cite{LYSO}) scintillators where their remaining energy is absorbed. 
The scintillators are 3$\times$3$\times$8~cm$^3$ blocks (with the corresponding dimensions in terms of radiation length 7$\times X_0$ and Moli\`ere radius 1.45$\times R_m$) arranged 
to form
a tight packing array with reflective
material (25~\micro m Teflon \cite{DuPont}) around each block and light-blocking material (50~\micro m Tedlar \cite{DuPont}) in between the blocks. 
The long axis of the blocks is parallel to the beam.

Readout of the scintillation light is accomplished using SiPM chips mounted directly 
onto
the rear end of the LYSO crystals and the edges of the plastic scintillator. 
This choice minimizes the space along the beam line devoted to the polarization measurement and reduces external fields that 
may
perturb any EDM measurement.

%
\subsection{LYSO Crystals}
\label{secLYSO}

\begin{figure}[hbtp] \centering
    \includegraphics[height=5cm]{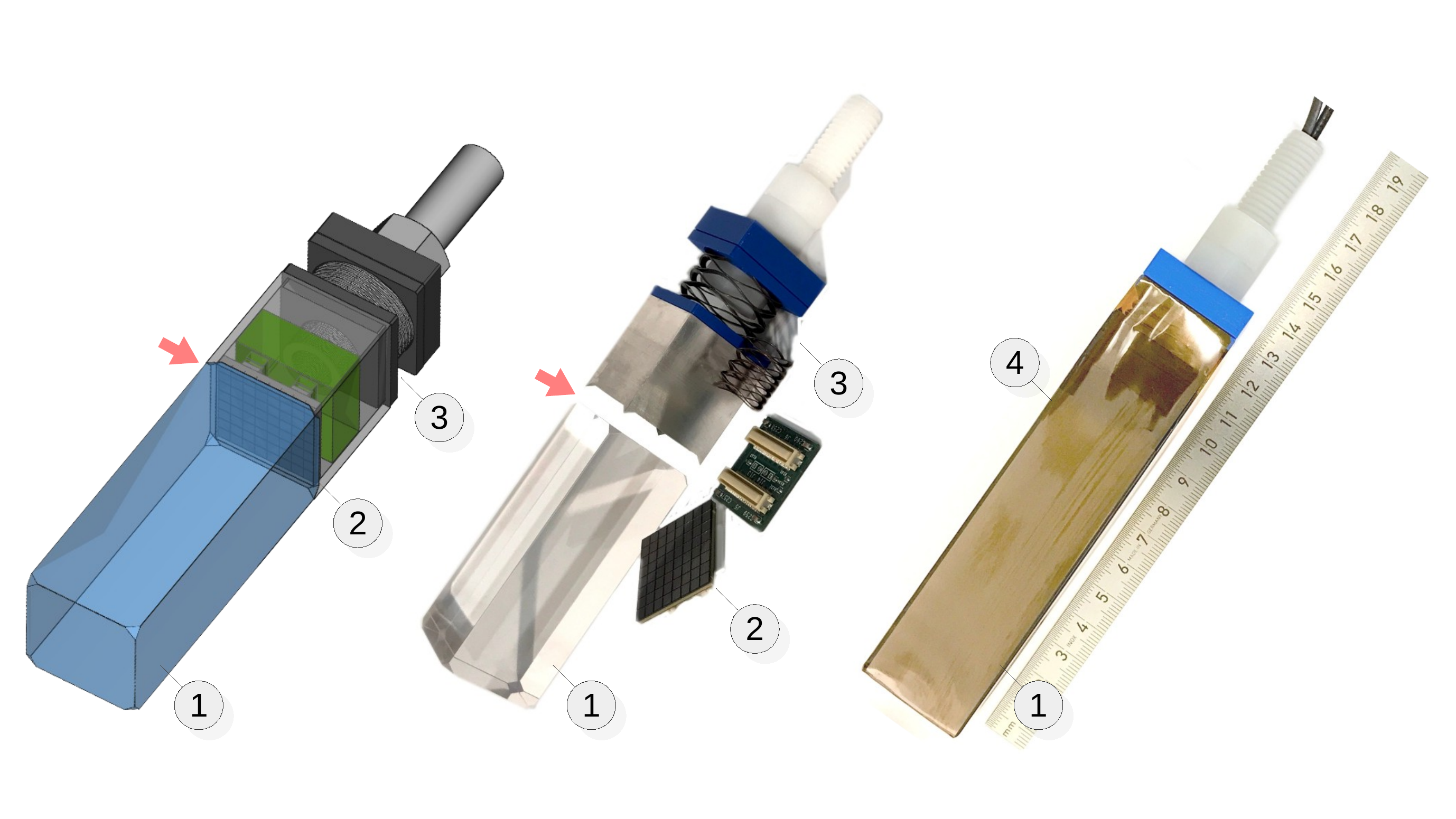}
    \caption{A full assembly sketch and disassembly photographs for a LYSO crystal detector with shading to indicate the main parts. 
    The red arrows indicate where the corners of the LYSO crystal are chamfered to
provide a stable mount against the aluminum housing that contains the SiPM. The labels indicate:
(1) a LYSO crystal; (2) SiPM array (8×8); (3) the housing and mechanical system; (4) wrapping with 3 layers: Teflon as a reflector, Tedlar to maintain light tightness, and Kapton \cite{DuPont} for mechanical stability.}
    \label{figLYSO}
\end{figure}

LYSO (cerium-doped lutetium yttrium oxyorthosilicate) crystals \cite{LYSO} are available commercially \cite{SG, EC, TPH} and have properties that make them nearly ideal
for an EDM calorimeter. 
The light output is high (32~photons/keV). 
The signal is fast (decay time about 41~ns at 420~nm, well matched to SiPM readouts). 
The crystals are not hydroscopic and they are dense (7.1~g/cm$^3$). 
As indicated in Fig. \ref{figSimFOM}, the current arrangement consists of 52 such crystals, 
48 of which are arranged
in 4×3 arrays along the four horizontal and vertical directions radiating out from the center. 
Along the axes, scattering angles between 4$^\circ$ and 15$^\circ$ are covered.
Fig.~\ref{figLYSO} shows a completed sketch of a module along with an open view of the readout end and the parts for another full detector.

The total length of the module is 14~cm. 
The crystal is coupled to a SiPM array via a 1~mm thick optical silicon interface \cite{SiliconPad}. 
The SiPM readout array is the SensL J-series with 20~\micro m pixels \cite{SensL}. 
The array's readout is made with a custom-made passive summarizer PCB, which adds up the charge from all 64 individual SiPMs.
First the crystal is wrapped in 25~\micro m white Teflon sheet, then it is wrapped in 25~\micro m black Tedlar sheet \cite{DuPont}.
The entire assembly is held together by two strips of 25~\micro m Kapton wrapped around the end of the LYSO crystal and pulled along the side, finally being held in place and tensioned with a circular wave-spring at the end. 
More details can be found in Refs.~\cite{FabianPhD, PSTP2017dito}. 

As the EDM search compares vertical polarization measurements across each beam store, stable operation of the detector gains and readout is essential.
The reverse bias voltage on the SiPM arrays is a compromise between maximizing overall gain and resolution and minimizing leakage current. 
For these particular SiPM arrays, operating voltages between 26~V and 31~V give good results. 
The routine operation is done with the 27~V supply voltage. 
This minimizes the power dissipation, and the signal height of the whole array reaches about 1~V. 
In addition, three modules are equipped with a digital thermometer.
The internal module temperature and the tunnel's outer temperature are continuously monitored and show an almost constant 1$^\circ$C difference.
This effect does not influence elastic peak, which is permanently monitored in online analysis, showing a stable gain factor depending on the tunnel temperature (which itself is stabilized within fractions of a degree).
A custom, 64-channel power supply \cite{Javakhishvili} was built for biasing the arrays. 
A 0.5~ppm/K ~reference voltage controls the regulation, yielding less than 0.02~\% overall instability. 
This is much less than the light collection resolution of the systems. 
In addition, a 128-channel monitoring and archiving system including on/off controls is used to record operating
conditions during each experiment.

%
\subsection{Data Acquisition and Readout}
\label{secDAQ}

\begin{figure}[hbtp] \centering
    \includegraphics[height=7cm]{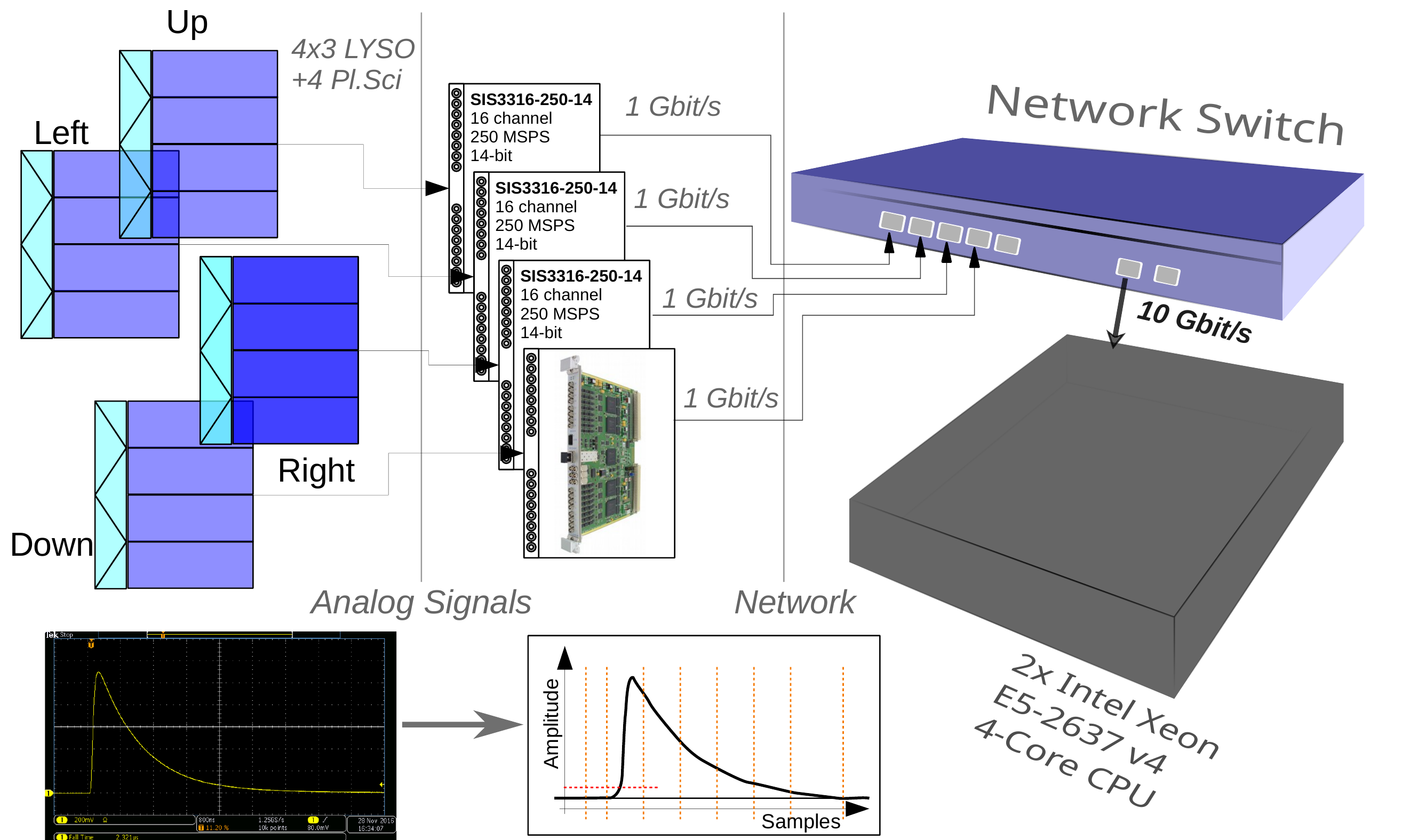}
    \caption{Diagram of the data acquisition system for the EDM polarimeter.}
    \label{figDAQ}
\end{figure}

The diagram of the data acquisition system is shown in Fig.~\ref{figDAQ}.
The readout is based on a “deadtimeless” design using high-resolution sampling ADCs based on a 250~MSPS, 14-bit ADC \cite{Struck}. 
Each pulse is divided into eight sampling regions as shown in the bottom of Fig.~\ref{figDAQ}. 
%
The first two of these, which precede the main pulse providing values for pedestal subtraction, and the other six are used to extract the amplitude. 
The FADC firmware provides a pile-up detection algorithm which sets the appropriate data bit. 
The signal integral distribution can also be checked for evidence of pile-up in the measured signal.
Additionally, in the online/offline analysis, the first two sample integration regions are checked for evidence of pile-up (if not equal) in the preceding signal.
A clock signal synchronized among all modules provides the basis for a time-stamping system that allows the time origin of the event to be located relative to the COSY machine RF. 
Events are recorded in a program based on a Linux server capable of providing both online and offline analysis. 
Only the values of the eight integrals are transferred to the main processor and stored.

%
\subsection{Typical Spectra}
\label{secSpectra}

\begin{figure}[hbtp] \centering
    \includegraphics[width=\textwidth]{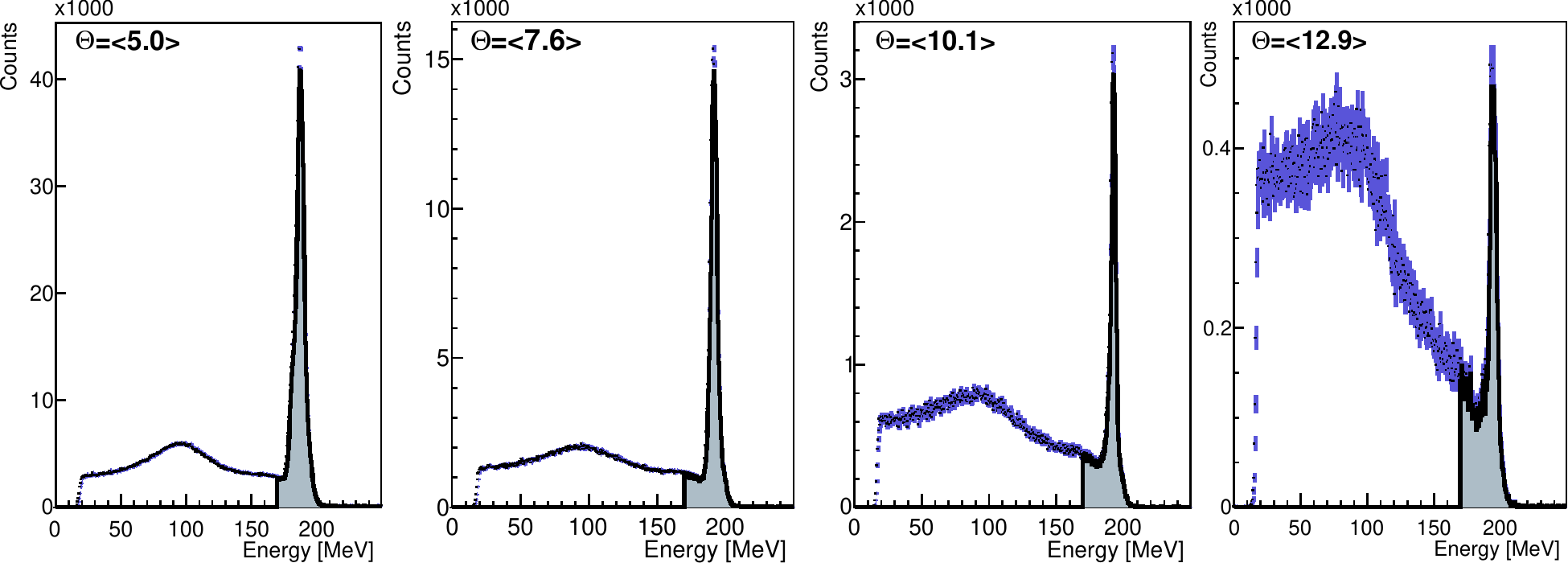}
    \caption{Pulse height spectra for one row of LYSO crystals shown in order of increasing scattering angle. The filled area indicates the fixed hardware threshold used during the event trigger. The cross section weighted scattering angle is also given for each spectrum.}
    \label{fig4_1D}
\end{figure}

\begin{figure}[hbtp] \centering
    \includegraphics[height=4.8cm]{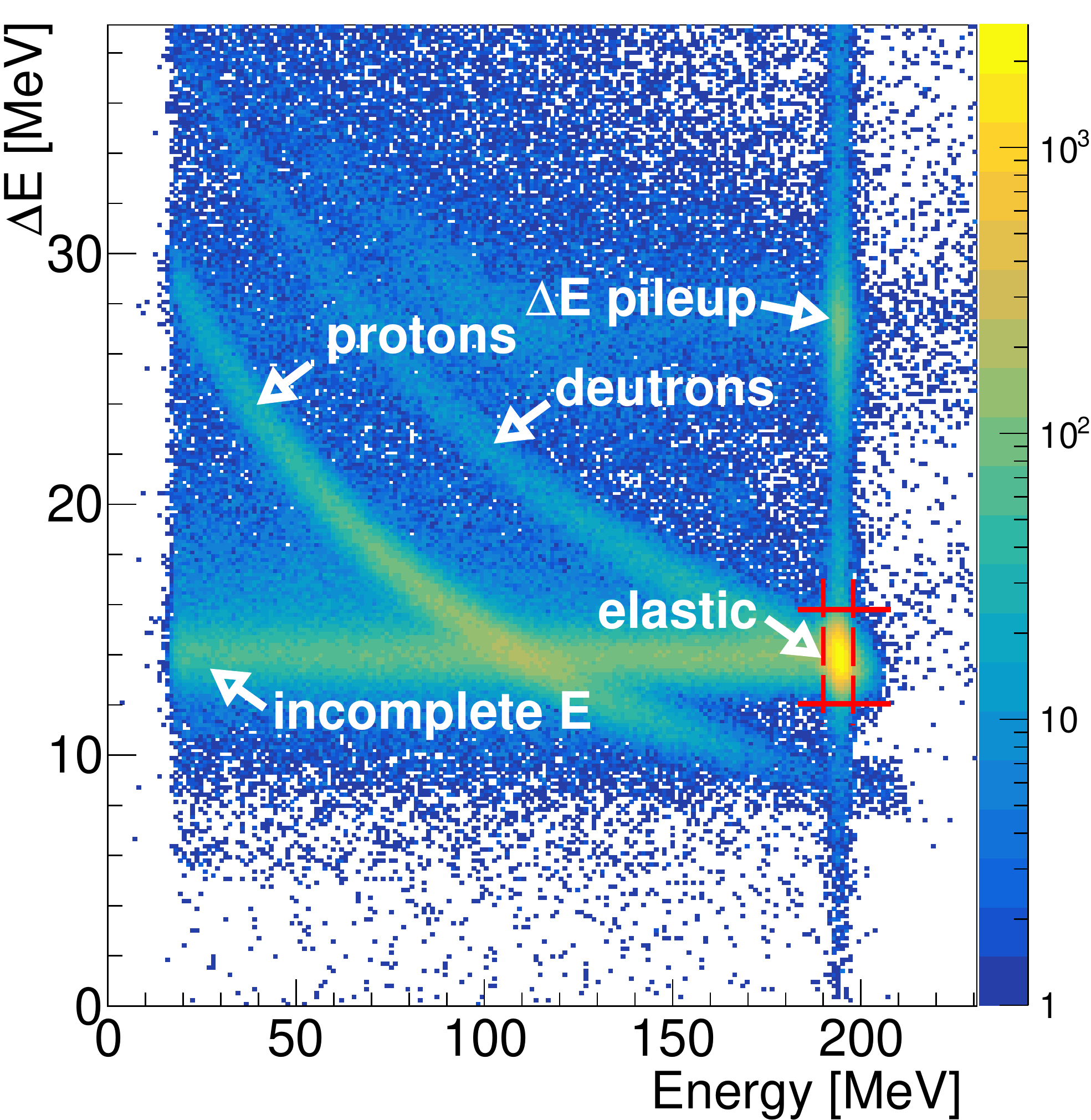}
    \hfill
    \includegraphics[height=4.8cm]{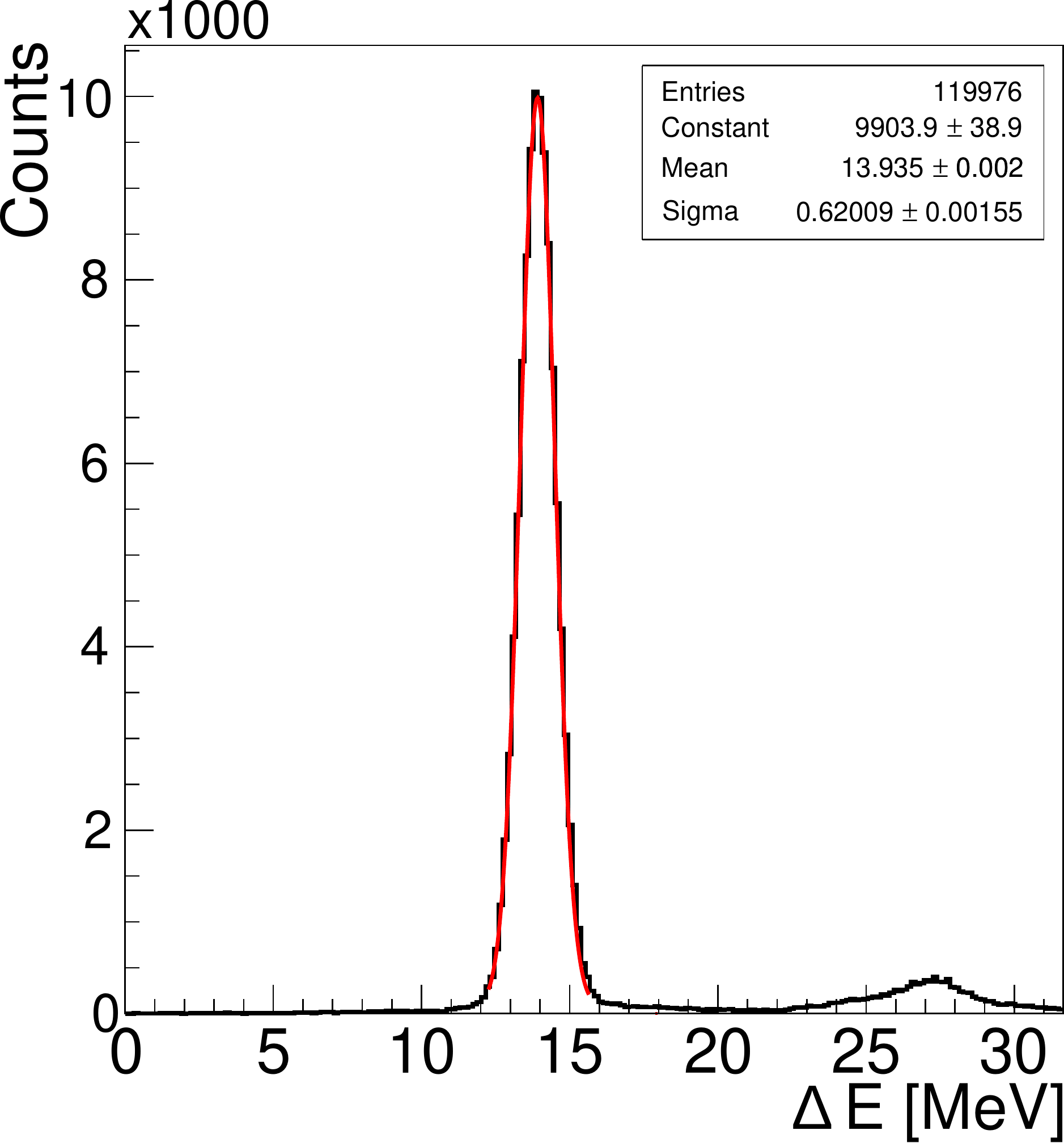}
    \hfill
    \includegraphics[height=4.8cm]{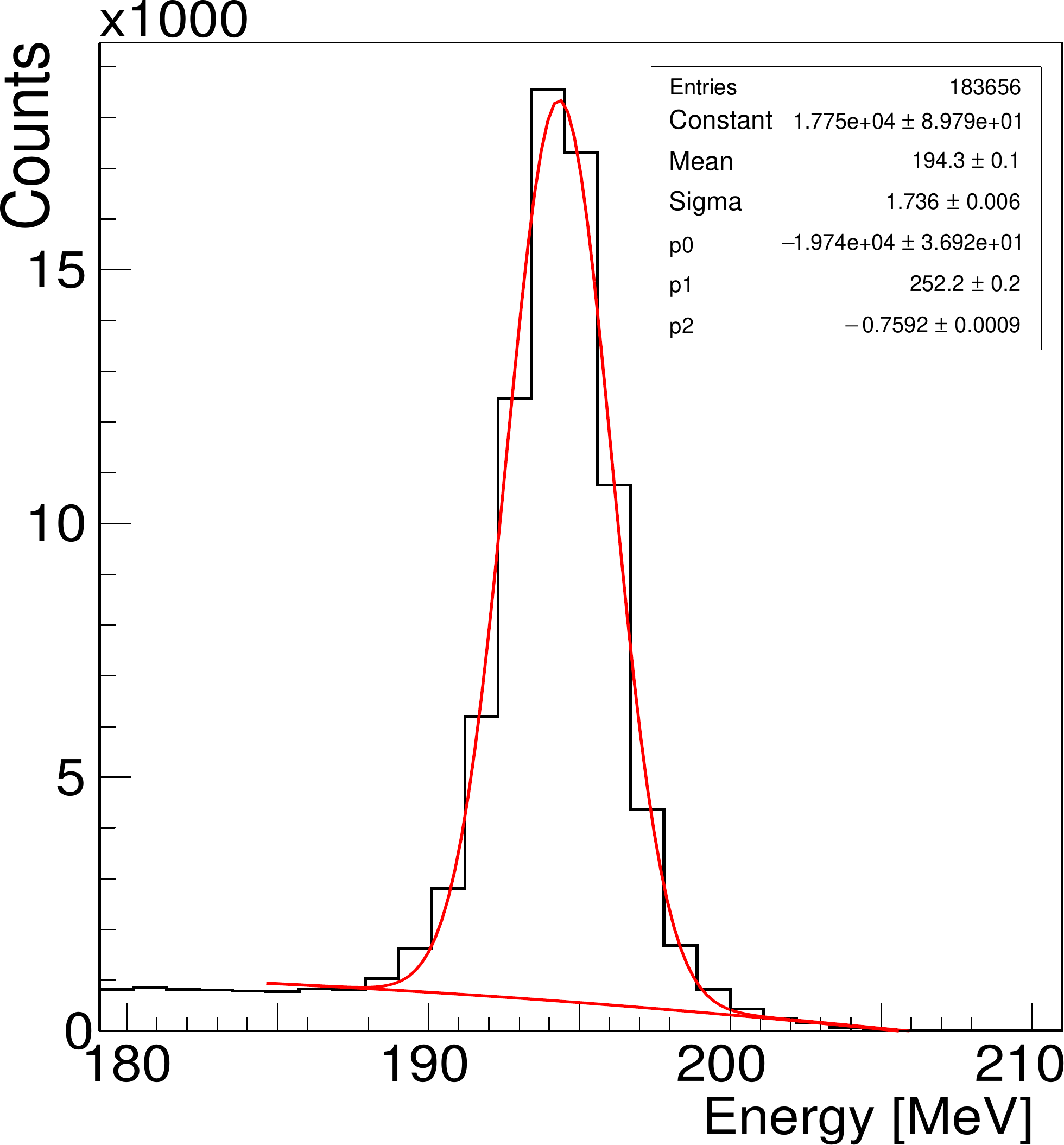}
    \caption{%
    (left panel) 2-D histogram of events as a function of LYSO pulse height on the X-axis
    and energy-loss ($\Delta E$) on the Y-axis. 
    The vertical limits marked in dashed red determine which range ($\pm3\sigma$) is used for the elastic projection in the middle panel. 
    The horizontal limits marked in red determine which events are recorded for further analysis as in the right panel and Fig. \ref{fig4_1D}. 
    (middle panel) Projection of the $\Delta E$ scintillator in the region of the elastic peak. 
    (right panel) Projection of the selected events on the LYSO crystal
    energy axis. A Gaussian shape with a linear background is used to illustrate the deuteron elastic peak. The sum for this
    feature without background subtraction is used as the primary data stream for determining polarization, as illustrated in Fig. \ref{fig4_1D}.}
    \label{fig2Dproj}
\end{figure}

Single-event 
spectra for a row of calorimeter detectors are shown in Fig.~\ref{fig4_1D}.
Each spectrum shows two features, a narrow peak at high pulse height that represents elastic deuteron
scattering from carbon and a broader peak at lower energy that is due mostly to protons from deuteron
breakup as well as a low-energy tail from the elastic peak. In addition to the LYSO crystals, there was also a general-purpose energy-loss scintillator mounted in front of the LYSO array. 
The resulting 2-D array is shown in Fig.~\ref{fig2Dproj}.
The most prominent peak in the left panel is from elastic scattering. 
Tails may be seen running parallel to both axes.
Events in which the deuteron interacts with the LYSO and loses energy make a line going to the left. 
Similar processes including pileup in the energy-loss scintillator account for the vertical band. 
A weak enhancement may be noted at double the energy-loss
value for elastic scattering due to two $\Delta E$ coincident events within the same time window. 
The curved arc starting from the elastic scattering peak represents inelastic deuteron reactions in the carbon target. 
A nearly parallel arc just below 
it
is due to protons. 
Midway along this arc is the main group representing protons from deuteron breakup in the carbon target.

For normal operation, a window may be placed around the deuteron peak in this spectrum in
order to generate a cleaner deuteron pulse height spectrum for actual polarimeter peak summing. 
Such a spectrum is shown in the right-hand panel of Fig.~\ref{fig2Dproj}. 
This peak sum is the primary piece of data used to determine the polarization. 
During normal operation, all events that pass the trigger threshold are (see Fig.~\ref{fig4_1D}) kept for the polarization measurement. 
Background subtraction as seen in the right-hand panel of Fig.~\ref{fig2Dproj} may be completed offline for more reproducible results.

\section{Installation}
\label{secInstallation}

\begin{figure}[hbtp] \centering
    \includegraphics[width=0.47\textwidth]{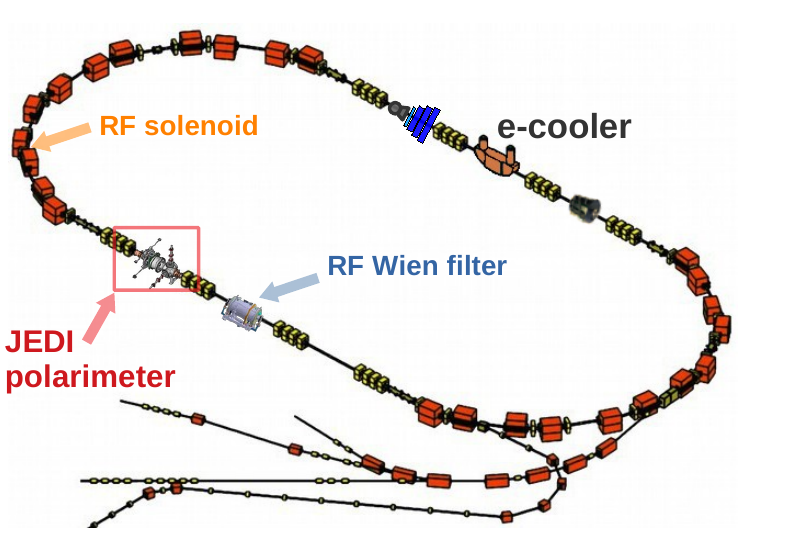}
    \hfill
    \includegraphics[width=0.47\textwidth]{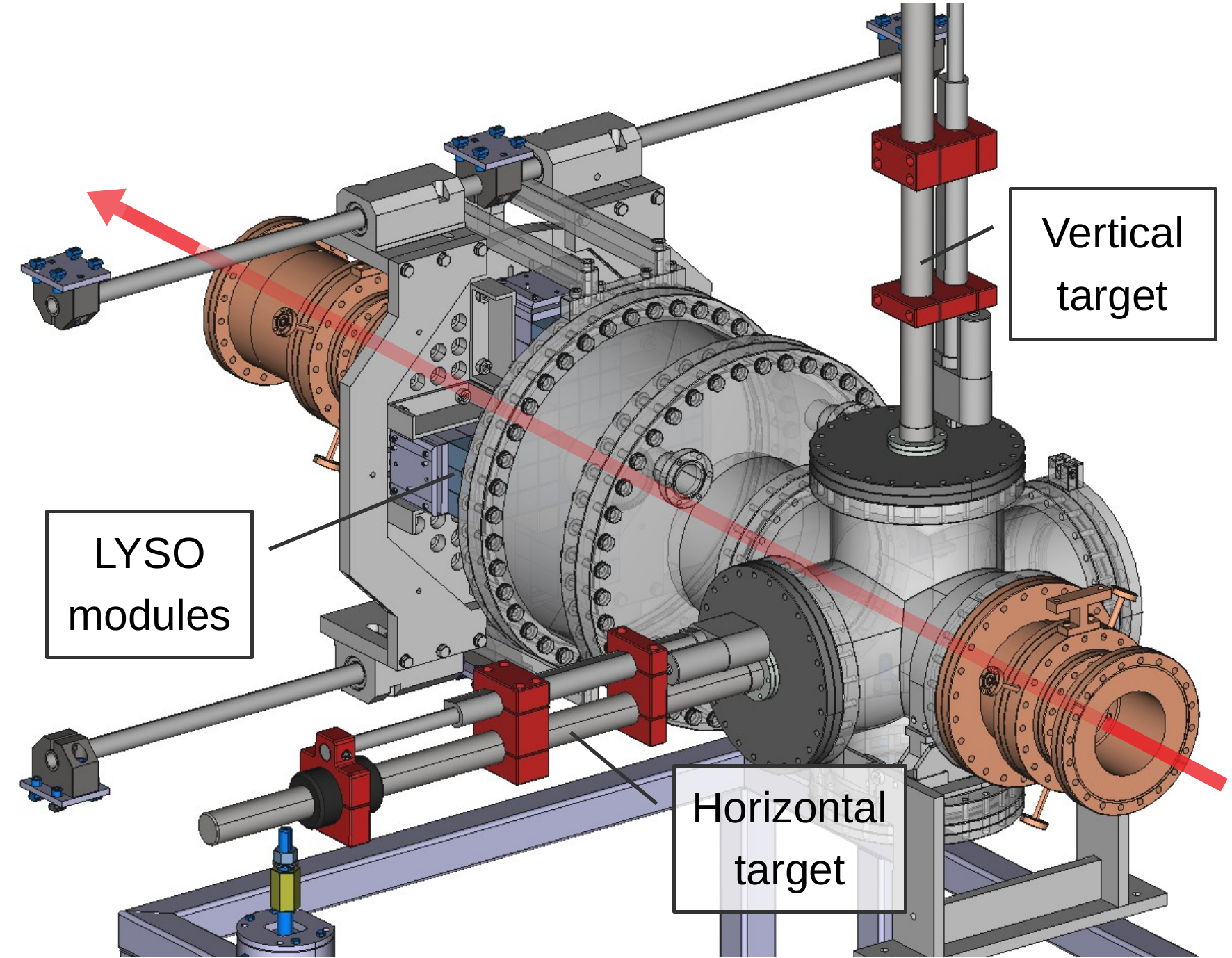}
    \caption{Left: The COSY storage ring (circumference 184~m) with a picture inset showing the location of the polarimeter installation. 
    The beam enters from the lower part of the diagram and circulates in a clockwise direction. 
    The additional beam pipe is for an extracted beam.
    Right: perspective drawing of the polarimeter prepared for installation. 
    The beam, indicated by the red arrow, enters from lower right. 
    The orange adapter flanges at the front and rear are planned to contain Rogowski coils to provide the X and Y beam position. 
    The 6-way cross contains two flanges that hold movable targets. 
    This is followed by an expansion chamber that ends with an 800~\micro m stainless steel exit window. Thus the detectors are in air. The LYSO (blue) crystals are
mounted on two movable tables. The energy loss ($\Delta E$) detectors are not included in this drawing, but would appear in front of
the LYSO calorimeter.}
    \label{fig45JEPO}
\end{figure}

At the beginning of 2019, the polarimeter was installed in the COSY ring for commissioning studies. 
The results are shown in the next section.
A perspective drawing of the polarimeter assembly is shown in Fig.~\ref{fig45JEPO}.

\begin{figure}[hbtp] \centering
    \includegraphics[height=5cm]{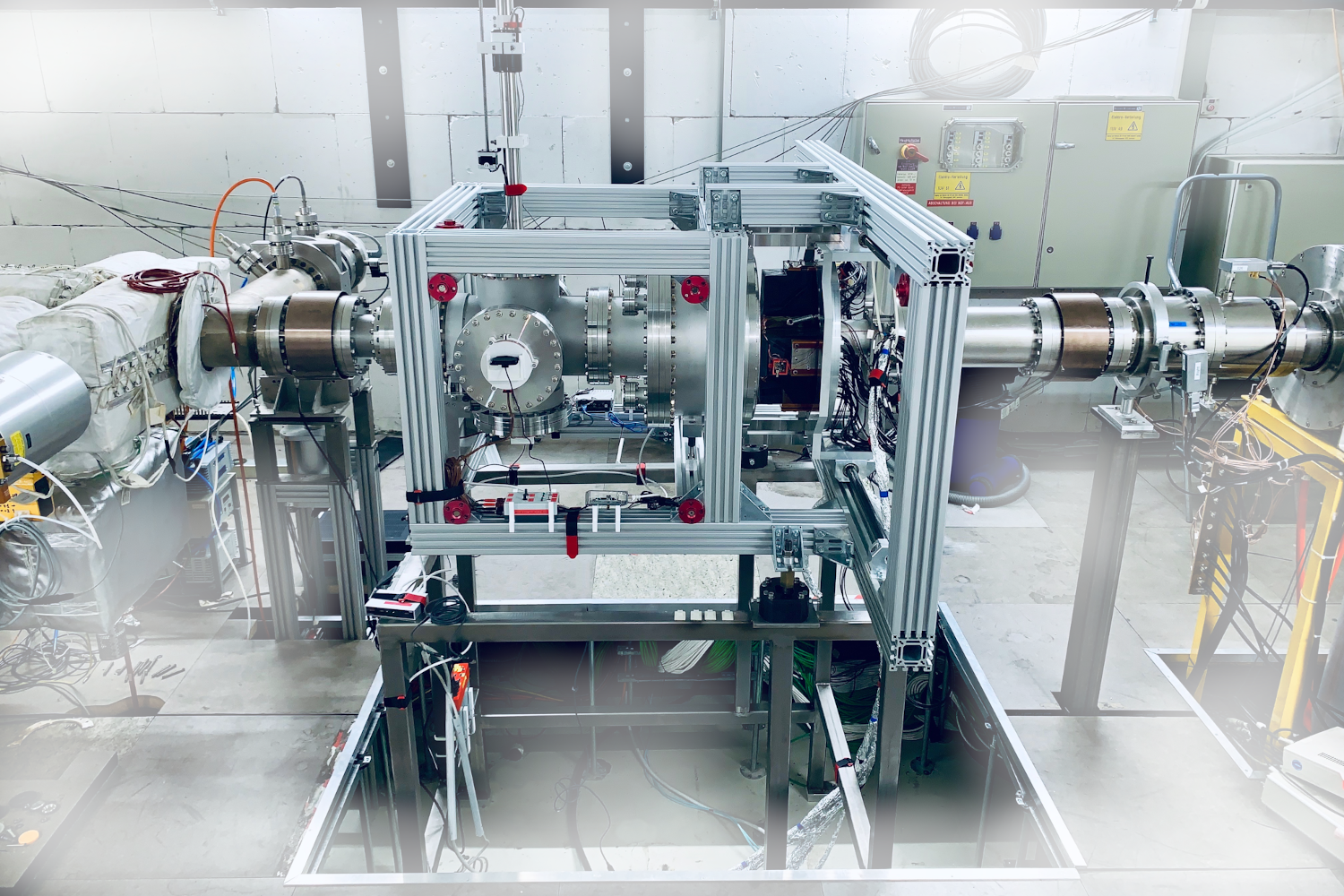}
    \hspace{1cm}
    \includegraphics[height=5cm]{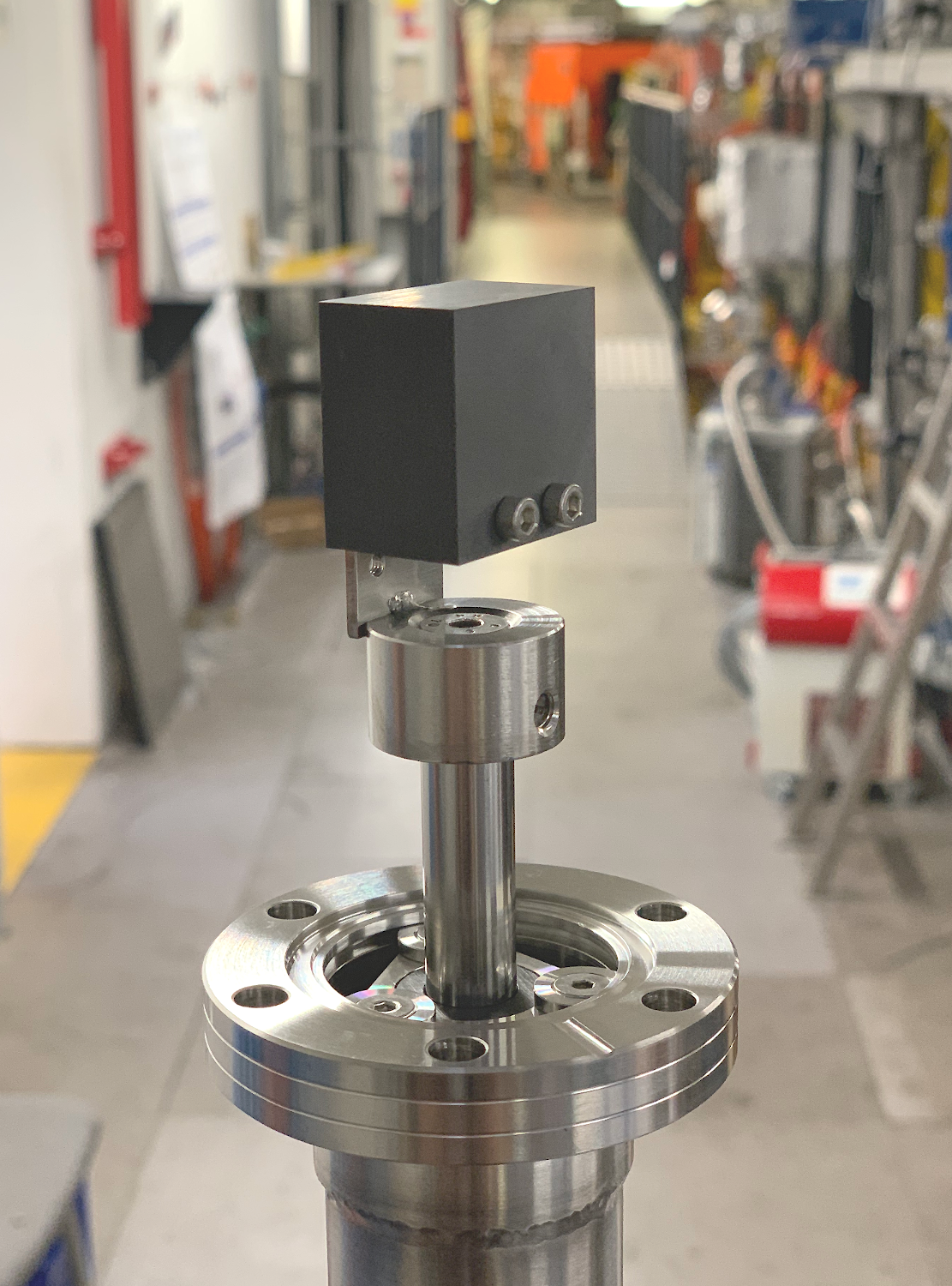}
    \caption{Left: The assembly of Fig.~\ref{fig45JEPO} in place in the COSY storage ring. The beam goes from left to
right.
Right: A graphite block target (2x3x3~cm$^3$) mounted on a movable rod. The beam would go about 3~mm above the top face of the carbon block.}
    \label{figJEPOpic}
\end{figure}

The main parts of the assembly are 
described
in the figure caption. 
The length of 1.48~m includes the space for Rogowski coil beam position monitors \cite{Rogowski}.
The assembly is mounted in a frame that 
fixes
it to the beam line, as shown in Fig.~\ref{figJEPOpic}.

The polarization is vertical at the point of injection, 
where the polarized negative ions are stripped and join the accumulating beam from the injector cyclotron. 
Devices along the beam are used for spin
manipulation, including the RF Wien filter and the RF Solenoid (located in the arc just after the
polarimeter). 
The electron cooling system (e-cooler) is used to reduce the phase space of the beam. 
It contains a
solenoid field to guide the electron beam, and the spin rotation is compensated by two oppositely
wound solenoids on either side of the cooler region.

%
\section{Results}
\label{l_results}

The polarimeter with 52 LYSO modules was 
integrated into
the COSY ring. 
These modules were arranged into four groups on a circular plate and were laid out into four rows of three crystals along radii that paralleled the four axes. 
In addition, one crystal was added along the diagonals in the corner where the four groups met, leaving a 3×3 crystal space open in the middle (see Fig.~\ref{figSimFOM}).

The polarized deuteron beam was injected into the COSY ring and accelerated to 970~MeV/c or 238~MeV, an energy commonly used for other COSY polarized-beam experiments. 
Three polarization states were available on a cycling basis, vector polarization up and down as well as unpolarized. 
These were created at the atomic beam polarized ion source using either a weak field transition unit or a combination of two strong field transition units \cite{Haeberli}.

In the beginning of the COSY cycle the beam was electron cooled for 75~s. 
Afterwards, the polarimeter used a 2~cm thick carbon (1.7~g/cm$^3$) block target that was inserted from above until the bottom face was aligned with the beam center line.
Prior to this, an orbit bump was created that moved the beam off-center at the target position. 
When data acquisition was enabled, white noise applied to a set of vertical strip line plates was used to increase the vertical emittance of the beam and bring deuterons close to the target where they could enter the front face and have an opportunity to scatter into the polarimeter detectors. 
The magnitude of the white noise was controlled in a feedback loop that maintained the counting rate in the polarimeter detector. 
The data acquisition phase operated for 120~s. 
Most of the beam in each store was eventually lost onto the polarimeter target.

%
\subsection{Calibration of the Beam Polarization}
\label{l_CalibBeamPol}

The transfer line between the JULIC cyclotron and COSY contains a scattering chamber used to determine the polarization of the beam. 
A thin carbon rod mounted at the center is the target. 
Around the outer edge are ports used for a collimation system and scintillation detectors.

At 40$^\circ$ the singles spectrum shows a clear elastic scattering peak that is easy to separate from other background reactions. 
Detectors were mounted at this angle in the left, right, up, and down directions. 
The left-right pair were used to measure the vector polarization. 
The asymmetry for each polarized state, given by $\epsilon = (L-R)/(L+R) = \frac{3}{2}p_y A_y$, was compared with the asymmetry for the unpolarized state. 
The difference was assumed to be due to the beam polarization. 
At this angle and energy, the elastic scattering analyzing power is $A_y = 0.61 \pm 0.04$ \cite{Chiladze}. 
The differences between the polarized and unpolarized state asymmetries for the two polarization states were $\epsilon (1) = -0.509 \pm 0.012$ and $\epsilon (2) = 0.178 \pm 0.012$, yielding polarizations of $p_y(1) = -0.550 \pm 0.013$ and $p_y(2) = 0.192 \pm 0.013$ with no contribution for the scale error of the calibration \cite{Chiladze}. 
The statistical errors given here are smaller than the systematic error associated with the calibration of the analyzing power.

%
\subsection{LYSO Polarimeter Spin Response}
\label{l_PolSpinResp}

\begin{figure}
    \centering
    \includegraphics[width=0.7\textwidth]{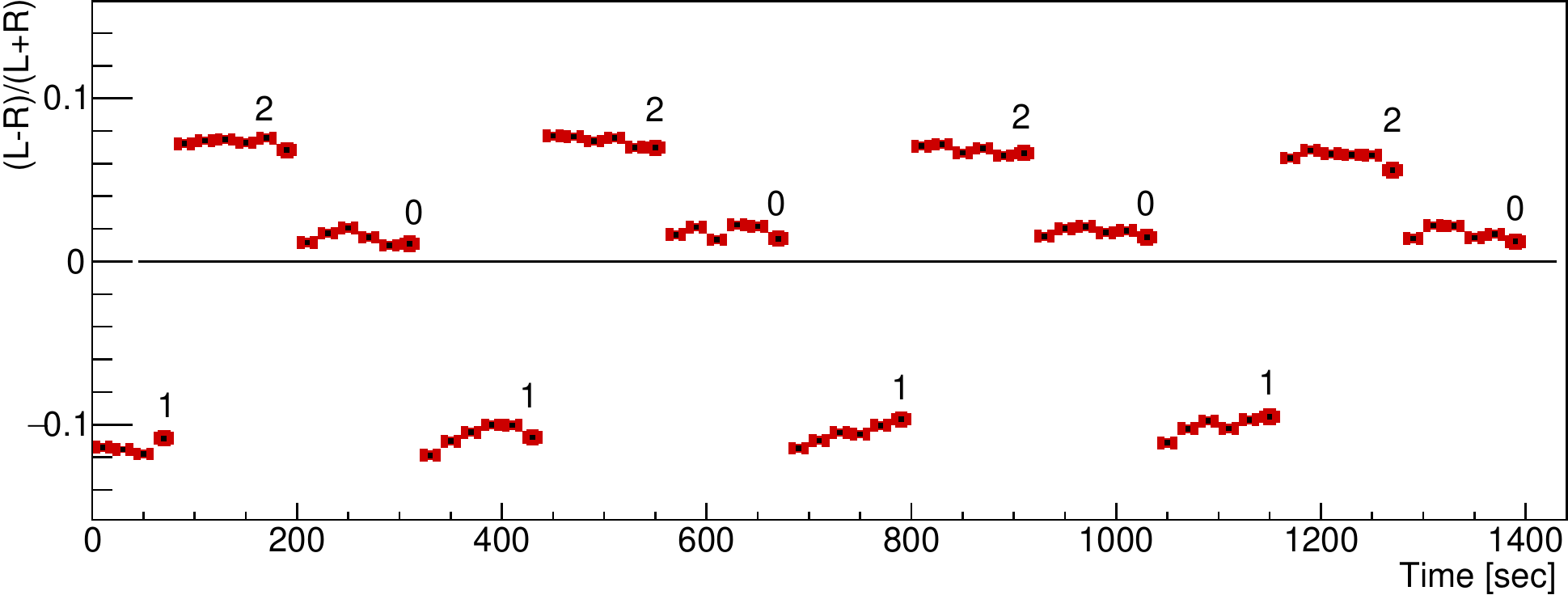} \\[3mm]
    \includegraphics[width=0.7\textwidth]{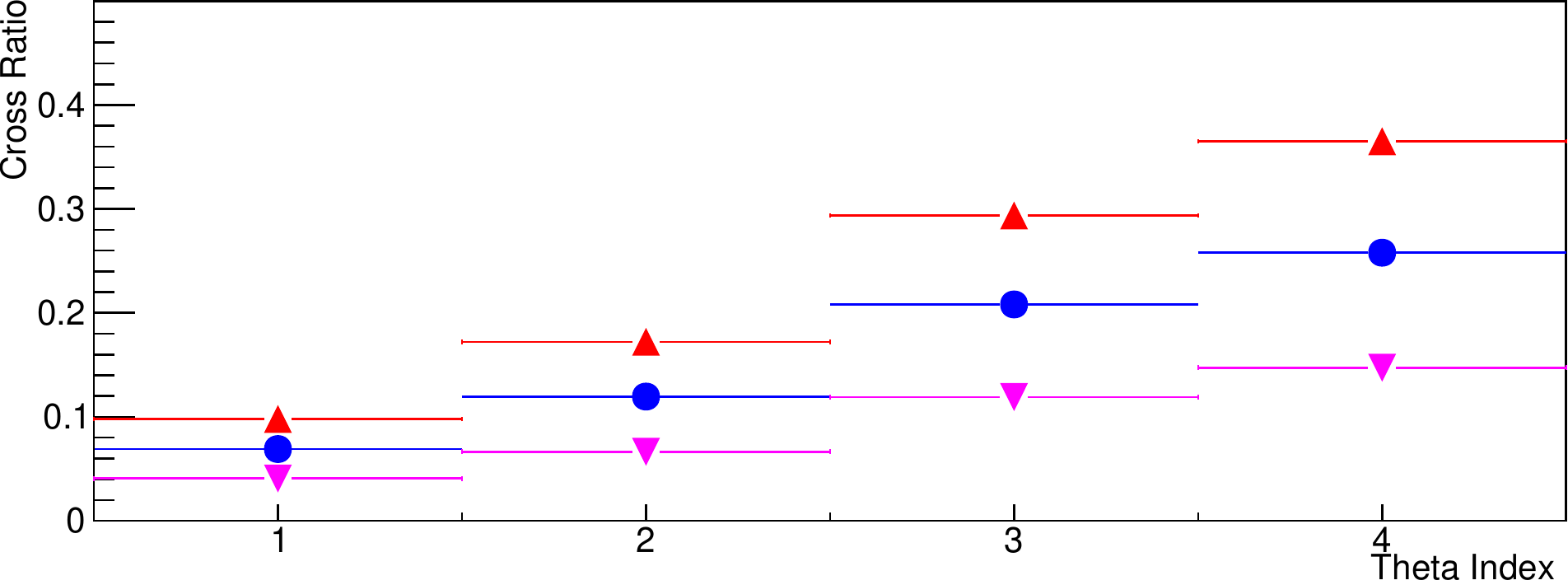}
    \caption{(Top) Raw asymmetries from a series of beam stores, in which the polarization cycle changes from negative (state 1) to positive (state 2) and then finally unpolarized (state 0). Only the data taking portion is retained. (Bottom) Asymmetries calculated for each column of three LYSO crystals in order from those closest to the beam to those farthest away. The 
    asymmetry values depicted in blue were calculated
    using the cross-ratio formula, Eq.~\ref{eqCR}, applied to the two polarized beam states. The red and magenta values represent half-cross-ratio (Eq.~\ref{eqHCR}) calculations based on the negative and positive polarization states, respectively, compared to the unpolarized state. Only the polarization magnitudes are shown.}
    \label{my_label}
\end{figure}

Figure \ref{my_label} illustrates the response of the polarimeter in two different ways.
The top panel shows the data acquisition portion of a series of fills. 
The polarization of the beam changes in a cyclical way.
The absolute value of the raw asymmetries decreasing within the cycle (120~s).
A small imbalance with the matching of left and right counting rates (due to either a steering or an alignment problem) results in all three asymmetries shifting upward by about 0.015. 
For this reason the average of the unpolarized measurements should be taken as the location of zero asymmetry.

The difference between the unpolarized and negative asymmetries may be used to determine the analyzing power of the polarimeter. 
The difference of -0.125 suggests a value of $A_y$(polarimeter) = 0.15 $\pm$ 0.01. 
This small value when summed over the entire left and right detector arrays reflects the heavy weighting given by the rates to the response of the first column of LYSO crystals by the small angle elastic scattering cross sections. 

\begin{figure}[hbtp] \centering
    \includegraphics[height=7cm]{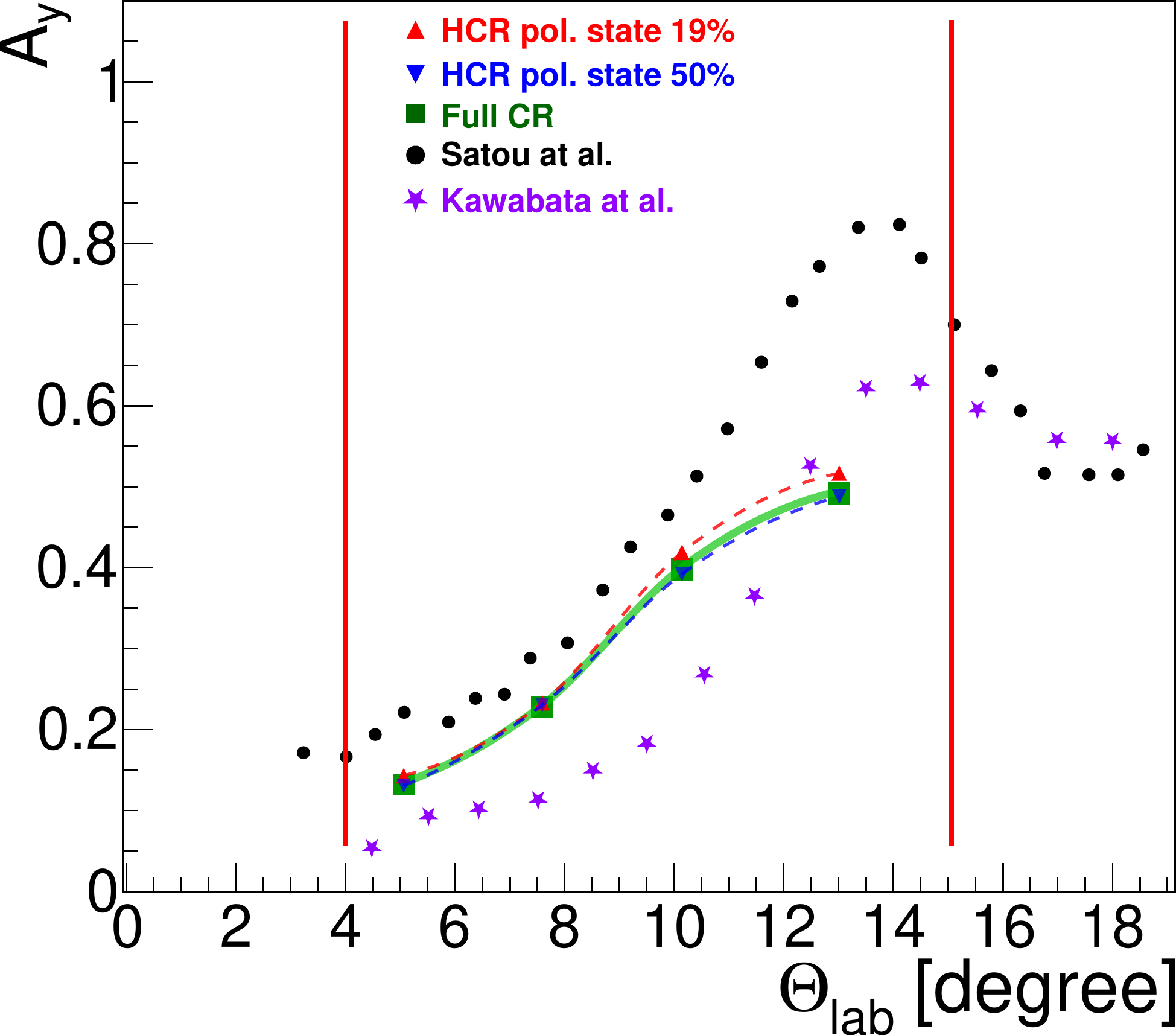}
    \caption{Values of the polarimeter analyzing power extracted for each column of the LYSO crystal array versus average laboratory scattering angle. Values are shown in blue and red for the ($-$) and (+) polarization states (half cross ratio) and in green for both polarized states together (cross ratio). Solid and dashed lines are guides to the eye. The angles for the four LYSO columns represent the cross-section weighted average scattering angle (see Fig.~\ref{fig4_1D}). For comparison, measurements for deuteron elastic scattering from carbon are shown at 270~MeV (black) \cite{Satou} and 200~MeV (blue) \cite{Kawabata}. The red lines mark the boundaries of the LYSO crystal array.}
    \label{figJEPO_CR}
\end{figure}

The values of the analyzing power are plotted as a function of average laboratory scattering angle in Fig.~\ref{figJEPO_CR}. 
The results are almost the same, regardless of the analysis method. 
The average angle (see Fig.~\ref{fig4_1D}) was computed with a weight proportional to the differential cross section. 
The two vertical red lines show the limits of the acceptance of the LYSO crystal array. 
As expected, this averaging process has moved the angles forward.

These results may be compared to the two measured angular distributions for elastic deuteron scattering closest in energy to our commissioning experiment. 
The analyzing power varies smoothly with energy, so we would expect that our polarimeter results at $T_d = $ 238~MeV would lay between the two sets of measurements at  $T_d = $ 200~MeV \cite{Kawabata} and  $T_d = $ 270~MeV \cite{Satou}. 
As seen in the figure, this is mostly the case, except for the result at the largest scattering angle. The reason for this is not fully understood, but it may be related to the relatively larger background contribution to the elastic scattering event rate, as seen in the rightmost panel of Fig.~\ref{fig4_1D}.

%
\subsection{Polarimeter Efficiency}
\label{l_PolEff}

The EDM search 
relies on the polarimeter being
efficient in extracting the most information about changes in the polarization. 
This efficiency is defined as the ratio of the number of events used to calculate the polarization to the number of beam particles consumed by an interaction with the polarimeter. 
Ideally, all the stored beam particles should interact with the polarimeter target.

During the experiment, the beam was bunched and the current monitored by a current transformer that surrounded the beam. 
Typical fills contained roughly 10$^9$ deuterons at the end of the electron-cooling time. 
The current transformer output was recorded continuously. 
In parallel, the count rate for the polarimeter detectors was also recorded. 

\begin{figure}[hbtp] \centering

    \includegraphics[height=8cm]{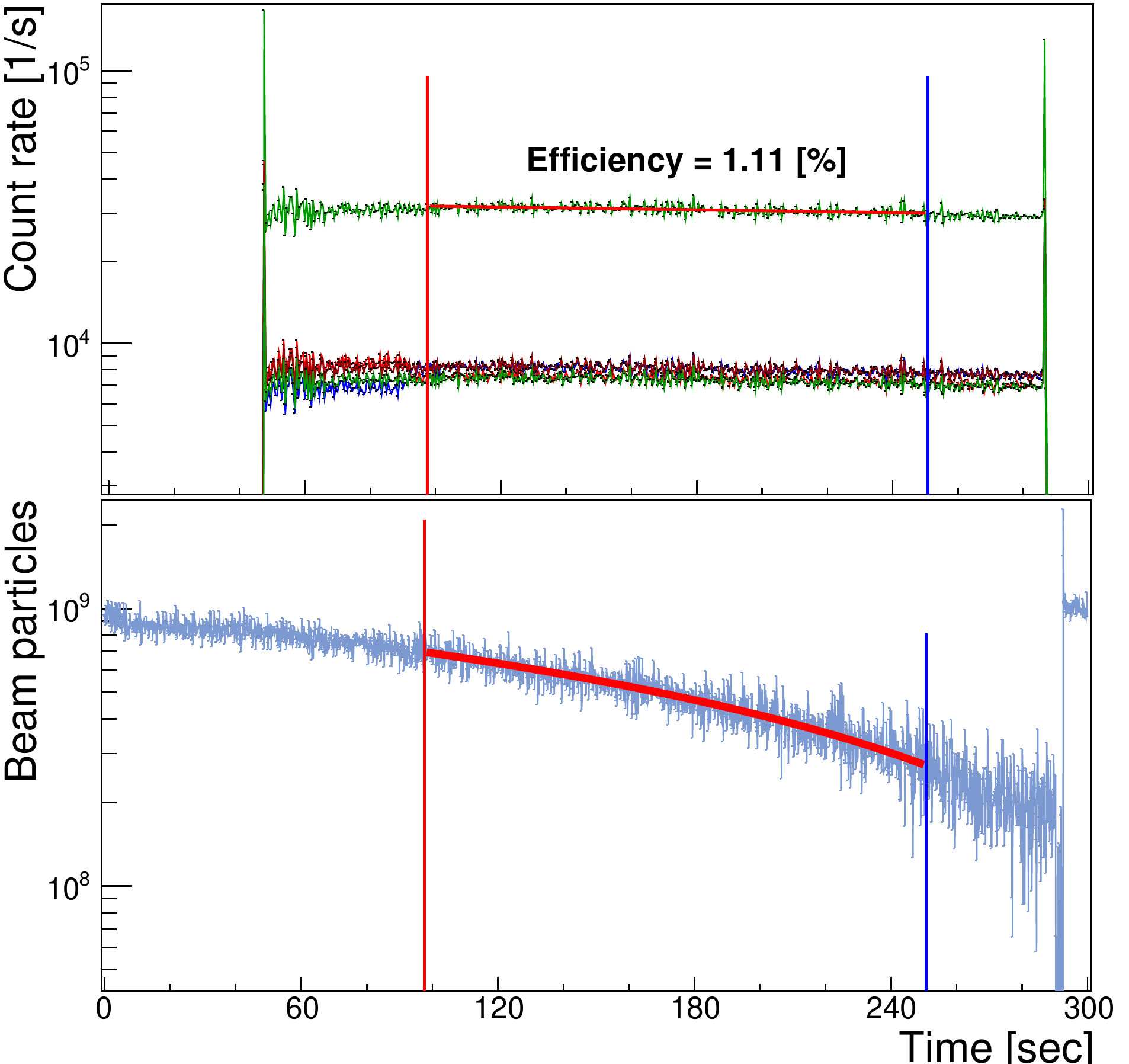}
    \caption{Part of the online logging system output for polarimeter runs. 
    The counting rates for the four quadrants of the polarimeter appear as colored traces on upper panel that are nearly flat for the data acquisition period. 
    The green trace shows the summed count rate.
    The start of beam extraction onto the target generates a spike in rate. 
    The blue trace in the lower panel shows the current monitor output of the declining stored beam intensity. 
    Times are shown along the horizontal axis in seconds.}
    \label{figPolEff}
\end{figure}

Sample data are shown in Fig.~\ref{figPolEff}.
The declining beam current is shown as a blue trace in the lower panel that slopes steadily downward after the start of data acquisition. 
This start time also applies to the white-noise generator that brings particles in the beam to the polarimeter carbon target. 
The count rates are regulated through a feed-back loop to the white noise generator, and they remain steady throughout the store. 
The sum of the count rates and the loss of the beam current may be read from this output for any desired time interval. 
The ratio of total events within a time interval marked by the red and blue vertical lines divided by the loss of beam particles for the same interval is about 1.1~\%. 
This value meets the requirement needed to satisfy Eq.~(3) of \cite{Yannis} (which considers the L/R counting rate only) for an EDM search on the proton at the level of $10^{-29}$~e$\cdot$cm.
\section{Conclusion and Outlook}
\label{l_conclusion}

We have reported on the design, construction, and commissioning of a polarimeter, whose properties lend 
themselves
to an application in a storage ring made for EDM searches. 
This device needs to be efficient with a high sensitivity to polarization. 
Its application to the EDM experiment also gains if the polarimeter 
is compact and does not generate stray fields large enough to perturb the EDM experiment itself. 
This polarimeter has been installed on the COSY storage ring at the  Forschungszentrum J\"ulich where it will be used for subsequent EDM development runs and experiments. 
The efficiency and analyzing power agree qualitatively with expectations based on deuteron elastic scattering properties as reported in the literature.

Future plans include movable absorbers mounted inside the target chamber that 
absorb breakup protons associated with deuteron beams and remove their contribution from the scintillator
signals. 
The $\Delta E$ scintillator will be divided into overlapping triangular sections that will be read out using individual SiPMs \cite{FabianPhD}. 
From this information, it should be possible to reconstruct the hit position on the $\Delta E$ detector. 
There will be two sets of bars, aligned horizontally and vertically.
Development is underway for a pellet target system in which the position of the pellet is tracked optically while it moves through the beam \cite{JKM}. 
There is also a plan to place Rogowski coils at either end of the polarimeter assembly to obtain a precise position and direction of the beam as it passes through the polarimeter \cite{Rogowski}.

%
\acknowledgments
The authors wish to acknowledge the support by the European Research Council via the ERC AdG srEDM (Contract number 694340);
by the EU Horizon 2020 research and innovation program, STRONG-2020 project, under grant agreement No. 824093;
and by a grant from the Shota Rustaveli National Science Foundation of the Republic of Georgia (SRNSF Grant No. 217854, "A first-ever measurement of the EDM of the deuteron at COSY"; SRNSF grant No 04/48, "Regional Doctoral Program in Theoretical and Experimental Particle Physics").
We also thank all involved members of the JEDI collaboration for their dedication and persistence towards this long-term project.


\end{document}